\documentclass[amsmath,amssymb,nofootinbib,notitlepage,superscriptaddress,twocolumn]{revtex4-2}

\usepackage{comment}
\usepackage{amsmath,amsfonts,amsthm,bm} 
\usepackage{graphicx}
\usepackage{dcolumn}
\usepackage{bm}
\usepackage{color}
\definecolor{RedWine}{rgb}{0.743,0,0}
\definecolor{RoyalBlue}{rgb}{0.25,.41,.88}
\definecolor{ForestGreen}{rgb}{.13,.54,.13}
\usepackage[backref,breaklinks,colorlinks,citecolor=ForestGreen]{hyperref}

\newcommand{\pd}{\partial}

\begin{document}

\preprint{APS/123-QED}

\title{Dynamics of ultrarelativistic charged particles with strong radiation reaction.\\ I. Aristotelian equilibrium state}

\author{Yangyang Cai}
\author{Samuel E. Gralla}
\affiliation{Department of Physics, University of Arizona, 1118 E 4th Street, U.S.A}
\author{Vasileios Paschalidis}
\affiliation{Department of Physics, University of Arizona, 1118 E 4th Street, U.S.A}
 \affiliation{Department of Astronomy, University of Arizona, 933 N Cherry Ave, U.S.A}

\begin{abstract}
Previous studies from the astrophysics and laser physics communities have identified an interesting phenomenon wherein ultrarelativistic charged particles experiencing strong radiation reaction tend to move along special directions fixed by the local electromagnetic field.  In the relativity literature these are known as the ``principal null directions'' (PNDs) of the Maxwell field.  A particle in this regime has ``Aristotelian'' dynamics in the sense that its velocity (rather than acceleration) is determined by the local field.  We study this Aristotelian equilibrium in detail, starting from the Landau-Lifshitz equation describing charged particle motion including radiation reaction.  Using a Frenet-Serret frame adapted to the PNDs, we derive the Lorentz factor describing motion along the local PND, together with drift velocities reflecting slower passage from one PND to another.  We derive conditions on the field configuration that are necessary for such an equilibrium to occur.  We demonstrate agreement of our analytic formulas with full numerical solutions of the Landau-Lifshitz equation in the appropriate regime.  
\end{abstract}

\maketitle

\section{Introduction}

The motion of a charged particle in an external electromagnetic field is a fundamental problem in electrodynamics.  Organized as a formal expansion in the particle charge $q$, the leading $O(q)$ force is the Lorentz force, and the sub-leading $O(q^2)$ force is the Abraham-Lorentz-Dirac self-force that describes radiation reaction \cite{Poisson:1999tv}.  The approximation is valid as long as the self-force is small compared with the Lorentz force in the particle rest frame, and the self-consistent motion is described by the Landau-Lifshitz (LL) equation \cite{LL,Poisson:1999tv,Gralla:2009md,DiPiazza:2011tq}.\footnote{Since quantum effects become important before this condition is violated \cite{LL,DiPiazza:2011tq}, the LL equation appears to be a complete description of the motion of classical point particles endowed with mass and charge, but no higher moments like spin and dipole.} However, as pointed out by LL in the original derivation, for ultrarelativistic particles, the \textit{lab-frame} self-force can be comparable to, or even much greater than, the lab-frame Lorentz force, while still respecting the basic validity of the equation.  We will call this the regime of \textit{strong radiation reaction}.\footnote{We avoid the term ``radiation dominated'' since typically the lab-frame Lorentz force is comparable to the radiation force.}

Interest in strong radiation reaction has grown in recent years in response to exciting experimental developments in two separate fields.  In laser-plasma physics, laser intensities are reaching the point where driven  particles will experience strong radiation reaction \cite{DiPiazza:2011tq}; and in astrophysics, unexpectedly bright gamma-rays from pulsars \cite{Fermi-LAT:2009orv,Fermi-LAT:2013svs} suggest that some pulsar magnetospheres may operate in this regime \cite{Gruzinov:2013pva,gruzinov2013b,Petri:2019tix,Cao:2019uhv,Cao:2021qrl,Petri:2022rym}.  Both communities have independently identified a rather interesting phenomenon wherein ultrarelativistic particles follow particular spatial directions determined entirely by the local electromagnetic field.\footnote{To our knowledge, this claim first appeared independently in \cite{Herold1985} and \cite{mestel1985}, before being rediscovered by \cite{Gruzinov:2013pva} and again by \cite{Gonoskov:2017lyz}.}  Since the electromagnetic field thus determines the \textit{velocity} of a particle, rather than its acceleration, we will follow \cite{mestel1985,Gruzinov:2013pva} in calling this regime ``Aristotelian''.

The Aristotelian velocities are most compactly characterized in an eigenvalue problem involving the field strength tensor $F_{\mu \nu}$ \cite{Jacobson:2015cia},
\begin{align}\label{eigen}
    F^{\mu}{}_{\nu} \ell^\nu = \lambda \ell^\mu,
\end{align}
where Greek letters denote spacetime indices, which are raised and lowered with the spacetime metric.  We may write the (real) solutions as 
\begin{align}
\lambda_\pm=\pm E_0, \qquad \ell^\mu = (1,\Vec{v}_\pm), \label{lambda} \\
\Vec{v}_{\pm}=\frac{\Vec{E}\times\Vec{B}\pm(B_0\Vec{B}+E_0\Vec{E})}{B^2+E_0^2}, \label{vpm}
\end{align} 
where $E_0$ and $B_0$ are given in terms of the invariants $P=\Vec{B}^2-\Vec{E}^2$ and $Q=\Vec{E}\cdot\Vec{B}$ as 
\begin{align}
    E_0 & = \sqrt{\sqrt{(P/2)^2+Q^2}-P/2} \label{E0} \\
    B_0 & = \textrm{sign}(Q) \sqrt{\sqrt{(P/2)^2+Q^2}+P/2}. \label{B0}
\end{align}
The eigenvectors $\ell_\pm^\mu$ are called the \textit{principal null directions} (PNDs) of the Maxwell field \cite{Penrose:1986ca}.

When $\Vec{E}\cdot\Vec{B} \neq 0$, we may always boost into a local frame where $\Vec{E}$ and $\Vec{B}$ are parallel, and this property is preserved under boosts in the field direction.  $E_0$ and $B_0$ are the electric and magnetic field strengths in this family of frames, with $E_0\geq 0$ by convention and the sign of $B_0$ reflecting whether $\Vec{B}$ is aligned or anti-aligned with $\Vec{E}$.  The velocities $\vec{v}_\pm$ are aligned with the fields in these special frames.  In the Aristotelian limit, particles move approximately along these directions, i.e., they have nearly light speed velocities given by
\begin{align}\label{eq:v=vpm}
    \Vec{v} \approx \Vec{v}_\pm,
\end{align}
with the plus/minus sign corresponding to positively/negatively charged particles.

Eq.~\eqref{eq:v=vpm} is the leading order description of the Aristotelian particle velocity.  However, the sub-leading velocity terms are important since they determine the leading Lorentz factor, which in turn determines the radiation emitted.  Gruzinov \cite{Gruzinov:2013pva} estimated the Lorentz factor $\gamma$ by reasoning that the power delivered by Lorentz force, $|q|E_0$, should be equal to the energy lost to curvature radiation, $(2/3) q^2 \gamma^4/R^2$, where $R$ is the radius of curvature of the trajectory.\footnote{A particle moving on a circular trajectory of radius $R$ emits power given by $(-2/3) q^2 \gamma^4/R^2$.}  One therefore concludes that $\gamma$ is determined as
\begin{align}\label{eq:gruzi}
    \gamma^4 \approx \frac{3}{2} \frac{E_0 R^2}{|q|}.
\end{align}

The evidence for this Aristotelian behavior consists of a mix of numerical and analytical arguments given by a number of different authors over the years \cite{Herold1985,BFink1989,Gruzinov:2013pva,Jacobson:2015cia,Gonoskov:2017lyz,Samsonov:2018skj,Ekman:2021vwg,Gonoskov:2021hwf,Samsonov2022}, which we review in App.~\ref{sec:previous} below.  These arguments are strongly suggestive, but not definitive or complete, and each presentation of the subject is wedded to the particular physical setting (pulsars or laser plasmas) under discussion. In particular, general conditions for validity of Aristotelian dynamics have not been established.  

In this paper we will study the Aristotelian equilibrium in detail and derive its complete description at sub-leading order in the velocity (which is leading order in the Lorentz factor).  We solve the LL equation using a Frenet-Serret frame attached to the principal null directions (integral curves of $\bm{v}_{\pm}$), which automatically includes subleading velocity corrections in a general and consistent manner.  We derive a general equation for the Lorentz factor $\gamma$, involving the torsion of the curve in addition to its curvature, which reduces to the formula \eqref{eq:gruzi} when the torsion can be neglected.  In performing the derivation, we identify general conditions on the local electromagnetic field that are necessary for the equilibrium to occur.  We demonstrate agreement of our analytic results with numerical solutions of the full LL equation in the appropriate regime.  A second paper will present more details of the numerical code, along with a more extensive exploration of the parameter space \cite{inprep}. 

Our work establishes conditions for the consistency of the Aristotelian regime and provides full details of the dynamics.  An important limitation is that we do not discuss the conditions under which entry into the regime will actually occur.  In fact, the most basic feature of the dynamics---particles following the principal null directions---is an \textit{assumption} of our derivation, not a result.  We are unable to prove that particles will enter the equilibrium under any given set of conditions, and we are unable to rule out the existence of other Aristotelian equilibria where the motion is not along the principal null directions.  We hope to address these issues in the future.

We work in Gaussian units with the speed of light set to unity.  Our spacetime metric is flat, with signature $(-+++)$, and our field strength tensor satisfies $F_{0i}=-E_i$, where $E_i$ is the electric field.

\section{Strong Radiation Reaction}\label{sec:LL}

A classical particle of charge $q$ and mass $m$ defines a length scale $\mathcal{R}$ and an electric field scale $\mathcal{E}$,
\begin{align}\label{scales}
    \mathcal{R} \equiv \frac{q^2}{m}, \qquad \mathcal{E} \equiv \frac{3}{2} \frac{m^2}{|q|^3}.
\end{align}
The length scale $\mathcal{R}$ is the particle size at which the electrostatic self-energy is of order the rest mass; for electrons, it is called the ``classical electron radius''.  The field scale $\mathcal{E}$ is the strength of the self-field at distance $\mathcal{R}$, with a factor of $3/2$ that we insert for later convenience.  Since we use units with the speed of light set equal to unity, $\mathcal{R}$ also defines a timescale, and $\mathcal{E}$ also defines a magnetic field scale.  We may anticipate that these are typical scales at which the classical point particle description will break down.  That is, one suspects that a necessary condition for the use of the classical approximation is that
\begin{align}
    \textrm{(rest frame typical field variation)} & \gg \mathcal{R} \label{condition1} \\
    \textrm{(rest frame typical field strength)} & \ll \mathcal{E}. \label{condition2}
\end{align}
In fact, quantum effects become important before these scales are reached (e.g., \cite{DiPiazza:2011tq}).\footnote{The Compton wavelength is a factor of $\alpha$ larger than $\mathcal{R}$, and the Schwinger field strength is a factor of $\alpha$ smaller than $\mathcal{E}$, where $\alpha$ is the fine structure constant.}

The motion of the particle through an external field $F_{\mu\nu}$ obeys the LL equation \cite{LL,Poisson:1999tv,DiPiazza:2011tq},
\begin{align}\label{LL}
     \frac{du^{\alpha}}{d\tau} & = \frac{q}{m} F^{\alpha}{}_{\beta}u^{\beta} +\frac{2}{3} \frac{q^3}{m^2}u^{\beta}\frac{d}{d\tau}F^{\alpha}{}_{\beta}+\frac{2}{3} \frac{q^4}{m^3}F^{\alpha}{}_{\sigma}F^{\sigma}{}_{\delta}u^{\delta} \nonumber \\
     &-\frac{2}{3} \frac{q^4}{m^3}(F^\beta{}_\gamma u^\gamma)(F_{\beta \delta} u^\delta)u^{\alpha},
\end{align}
where $u^\alpha$ is the four-velocity and $\tau$ is the proper time.  This first term on the right-hand side is the Lorentz force, and the remaining three comprise the self-force.  This equation is derived under the assumption that the self-force is small compared to the Lorentz force in the particle rest frame \cite{LL,Gralla:2009md,DiPiazza:2011tq}.  To understand the meaning of this condition, let us first write the equation in terms of the fundamental scales \eqref{scales},
\begin{align}\label{LL2}
     \mathcal{R} \frac{du^{\alpha}}{d\tau} & = \pm\frac{1}{\mathcal{E}} F^{\alpha}{}_{\beta}u^{\beta} \pm\frac{\mathcal{R}}{\mathcal{E}}u^{\beta}\frac{d}{d\tau}F^{\alpha}{}_{\sigma}+\frac{1}{\mathcal{E}^2}F^{\alpha}{}_{\sigma}F^{\sigma}{}_{\delta}u^{\delta} \nonumber \\
     &-\frac{1}{\mathcal{E}^2}(F^\beta{}_\gamma u^\gamma)(F_{\beta \delta} u^\delta)u^{\alpha}.
\end{align}
where $\pm$ is the sign of $q$.  
Viewing this equation in a frame where the particle is instantaneously at rest, the four-velocity components are order unity.  It is then evident that the conditions for validity of the LL equation (last three terms on right-hand side small compared to first term) are equivalent to the general conditions \eqref{condition1} and \eqref{condition2}.

Now suppose that we work in a global ``lab frame,'' and let $F$ denote a typical field strength in this frame. The rest frame field strength can be up to a factor of $\gamma$ times larger, in which case the condition \eqref{condition2} becomes
\begin{align}\label{condition2-2}
    \frac{\gamma F}{\mathcal{E}} \ll 1,
\end{align}
which in particular implies $F/\mathcal{E} \ll 1$.  In the lab frame, each occurrence of $u^\alpha$ introduces a factor of $\gamma$, which can be large.  The last term in \eqref{condition2-2} involves three factors of $\gamma$, and hence can be as large as $\gamma^3 F^2/\mathcal{E}^2$.  Comparing to the first term, the naive ratio of self-force to Lorentz force is
\begin{align}\label{naive}
    \frac{\textrm{(lab-frame self-force)}}{\textrm{(lab-frame Lorentz force)}} \lesssim \gamma^2 \frac{F}{\mathcal{E}}.
\end{align}
This ratio can be arbitrarily large while still respecting the basic condition \eqref{condition2-2}.  It  is order unity for the Aristotelian equilibrium, due to special cancellations from combinations of tensor components.

The covariant formulation \eqref{LL} is elegant, but inconvenient for studying the Aristotelian limit.  For example, since the motion is nearly along the PND $\ell^\mu$, it seems natural to write $u^\mu \approx \gamma \ell^\mu$ with $\gamma \gg 1$.  However, this expression leads to contradictions, since it implies $u^\mu u_\mu = 0$ instead of $u^\mu u_\mu = -1$.  Instead, the Aristotelian limit is more easily studied as an expansion in the three-velocity, $\Vec{v} \approx \Vec{v}_\pm + \delta \Vec{v}$, where the small subleading corrections $\delta \Vec{v}$ will encode the large Lorentz factor.  To perform this expansion we will introduce a Frenet-Serret frame adapted to the PND, as follows.

\section{Frenet-Serret Frame}\label{sec:FS}

The principal null directions $\ell^\mu_\pm = (1,\Vec{v}_\pm)$ [see Eqs.~\eqref{lambda} and \eqref{vpm}] are defined at every point in space by the local electromagnetic field.  The spacetime curves that are tangent to the principal null directions define two sets of lightlike trajectories filling all of space---the principal null \textit{congruences}, in relativity terminology.  In a given global inertial frame (the lab frame), each such trajectory may be described as $\vec{x}_{\pm}(t)$, where
\begin{align}
    \frac{d \Vec{x}_{\pm}}{d t} = \Vec{v}_\pm(\Vec{x},t).
\end{align}
Since $v_\pm^2=1$, this is actually an arclength parameterization of the space curve $\Vec{x}_{\pm}(t)$, which is naturally suited to the Frenet-Serret formalism. we will denote the frame as ($\Vec{l},\Vec{n},\Vec{k}$) with
 \begin{align}\label{dl}
     \Vec{l} & = \Vec{v}_{\pm} \\ 
     \Vec{n} & = \frac{1}{\kappa} \frac{d\Vec{l}}{dt}\\
     \Vec{k} & =\Vec{l}\times\Vec{n}.
 \end{align}
The frame vectors ($\Vec{l},\Vec{n},\Vec{k}$) are unit vectors, and $\kappa$ is the curvature of the curve.  We also have the Frenet-Serret relations
\begin{align}
    \frac{d\Vec{n}}{dt} & =-\kappa\Vec{l}+\iota\Vec{k}\label{dn}\\
    \frac{d\Vec{k}}{dt} & =-\iota\Vec{n}\label{dk},
\end{align}
where $\iota$ is called the torsion of the curve.  

We denote the frame components of the electric and magnetic fields as follows,
\begin{align}
    \Vec{E}&=e_l\Vec{l}+e_n\Vec{n}+e_k\Vec{k}\\
    \Vec{B}&=b_l\Vec{l}+b_n\Vec{n}+b_k\Vec{k}.
\end{align}
The eigenvalue equation \eqref{eigen} becomes
\begin{align}
    \Vec{E}+\Vec{l}\times\Vec{B}&=\pm E_0\Vec{l}\label{pnd1}\\
    \Vec{E}\cdot\Vec{l}&=\pm E_0\label{pnd2},
\end{align}
which implies
\begin{align}
    e_l&=\pm E_0\\
    b_l&=\pm B_0\\
    e_n&=b_k\\
    e_k&=-b_n.
\end{align}
We may also decompose the velocity of any particle in this frame,
\begin{align}\label{vdecomp}
\Vec{v}=v_l\Vec{l}+\Vec{v}_{\perp}=v_l\Vec{l}+v_n\Vec{n}+v_k\Vec{k}.
\end{align}
We will denote the magnitude of $\Vec{v}_{\perp}$ as
\begin{align}
    v_\perp = \sqrt{v_n^2 + v_k^2}.
\end{align}

\section{Aristotelian equilibrium}\label{sec:derivation}

In 3+1 form, the LL equation \eqref{LL2} becomes
\begin{widetext}
\begin{align}
    \frac{d\gamma}{dt}&=\frac{q}{m}\left\{ \Vec{E}\cdot\Vec{v}+\frac{2}{3}\gamma \mathcal{R}\frac{d\Vec{E}}{dt}\cdot\Vec{v}\pm\frac{1}{\mathcal{E}}(\Vec{E}+\Vec{v}\times\Vec{B})\cdot\Vec{E}\mp\frac{\gamma^2}{\mathcal{E}}\left[(\Vec{E}+\Vec{v}\times\Vec{B})^2-(\Vec{E}\cdot\Vec{v})^2\right]\right\}\label{f0}\\
    \frac{d\Vec{v}}{dt}&=\frac{q}{m\gamma}\left\{ \vec{F}+\frac{2}{3}\gamma\mathcal{R}\vec{F}'\pm\frac{1}{\mathcal{E}}\left[(\Vec{v} \times \Vec{E} ) \times \Vec{E} + (\Vec{v}\times\Vec{B}) \times \Vec{B} + \Vec{E} \times \Vec{B} +\Vec{v} \cdot (\Vec{E} \times \Vec{B})\vec{v}\right]\right\}, \label{fi}
\end{align}
\end{widetext}

where we define 
\begin{align}\label{F}
    \Vec{F} = \vec{E} + \vec{v} \times \Vec{B} - (\vec{E} \cdot \vec{v})\Vec{v}
\end{align}
as well as
\begin{align}\label{F'}
    \Vec{F}' = \frac{d\vec{E}}{dt} + \vec{v} \times \frac{d \Vec{B}}{dt} - \left(\frac{d\vec{E}}{dt} \cdot \vec{v}\right)\Vec{v}.
\end{align}
In these expressions, $d/dt$ indicates a total derivative, resolvable as $d/dt=\pd/\pd t + \Vec{v} \cdot \Vec{\nabla}$ when acting on fields.  In Eqs.~\eqref{f0} and \eqref{fi}, the $\pm$ is the sign of $q$ and also labels the PND $v_\pm$ on which charges of that sign will move.

Our strategy will be to identify conditions under which the Aristotelian equilibrium occurs and use those conditions to derive its detailed properties.  Our assumptions will refer to a given ``lab frame''---see Sec.~\ref{sec:conditions} for further discussion. We seek a situation where the self-force is comparable to the Lorentz force, and  will drop all self-force terms that are much smaller than the Lorentz force.

\subsection{Assumptions}\label{sec:assumptions}

We begin with the scale of variation of the field.  If $L$ and $T$ are typical length and time scales, then the basic condition  \eqref{condition1} requires $L/\gamma \gg \mathcal{R}$ and $T/\gamma \gg \mathcal{R}$. We will also need to assume that the configuration is quasi-stationary in the sense that $T \gg L$.  Our assumptions on the scale of variation are thus
\begin{align}\label{nofieldderiv}
    T \gg L \gg \mathcal{R} \gamma,
\end{align}

Next consider the magnitude of the field.  The general condition \eqref{condition2} requires that the rest frame field strength be small compared to the critical field strength.  However, for Aristotelian equilibrium the lab and rest frame field strengths agree: the particle moves along a PND, so the relevant boost is along the field direction, which leaves the fields invariant. Thus we assume 
\begin{align}\label{weakfield}
    \{E,B\} \ll \mathcal{E},
\end{align}
where $E$ and $B$ are the lab-frame field strengths.  (The stronger condition $\gamma\{E,B\} \ll \mathcal{E}$ would guarantee \eqref{condition2} in all circumstances, but we will only need \eqref{weakfield} for the validity of Aristotelian equilibrium.) Note that \eqref{weakfield} also implies that the invariants $E_0$ and $B_0$ are small compared with $\mathcal{E}$.

Next we assume explicitly that the motion is primarily along a PND.  In the notation of Sec.~\ref{sec:FS}, we have
\begin{align}\label{approx}
    \gamma \gg 1, \qquad v_\perp \ll 1.
\end{align}
The first condition guarantees ultrarelativistic motion, while the second ensures that the spatial direction coincides with the PND.

Finally we assume that the particle is locally in equilibrium, with the leading-order power delivered by the Lorentz force balancing the leading-order power lost to the self-force.  This means that the energy of the particle changes slowly compared to the timescale set by either force.  Since the Lorentz force power is $q \vec{E} \cdot \Vec{v}_{\pm}=\pm q E_0$ at leading order, our assumption is
\begin{align}\label{gammadot}
    m\left|\frac{d\gamma}{dt}\right| \ll |q| E_0.
\end{align}

Eqs.~\eqref{nofieldderiv}, \eqref{weakfield}, \eqref{approx} and \eqref{gammadot} are the key physical assumptions underlying our derivation.  However, as we proceed through the calculations, we will see that these are not quite sufficient to derive local equilibrium conditions, since time derivatives of Frenet-Serret frame components would remain in the equations.  In order to be able to drop all time derivatives, we will be forced to assume the following four additional conditions:
\begin{align}
    \left|\frac{d v_l}{dt}\right| & \ll v_n  \kappa \label{dropvl} \\
    \left|\frac{d v_n}{dt}\right| & \ll v_l \kappa \label{dropvn} \\
        \left| \frac{dv_k}{dt}\right| & \ll \frac{|q| E_0}{\gamma m}\frac{1+(\gamma v_\perp)^2}{(\gamma v_\perp)^2}|v_k|. \label{dropvk1}
     \\
 \left|\frac{dv_k}{dt}\right| & \ll \frac{E_0}{|B_0+\frac{\gamma m}{|q|}\iota|}\frac{1+(\gamma v_\perp)^2}{(\gamma v_\perp)^2}\kappa \label{dropvk2}
\end{align}
However, once the equilibrium formulas for the velocity components are derived, it can be seen (appendix~\ref{sec:cleanup}) that these conditions are strictly weaker than those already imposed.  We discuss this issue further in Sec.~\ref{sec:conditions} below. 

\subsection{Derivation}

We now use the assumptions of Sec.~\ref{sec:assumptions} to derive the Aristotelian equilibrium from the LL equation \eqref{f0} and \eqref{fi}.  First note that Eq.~\eqref{gammadot} implies that the right-hand side (RHS) of Eq.~\eqref{f0} vanishes at leading order in our ultrarelativistic expansion, which by Eqs.~\eqref{nofieldderiv}, \eqref{weakfield} and \eqref{approx} becomes
\begin{align}\label{f0balance}
    E_0=\frac{\gamma^2}{\mathcal{E}}(E_0^2+B_0^2)v_{\perp}^2.
\end{align}
This equation comes entirely from the first and last terms of the RHS of \eqref{f0}.  (The second term is negligible compared to the first term by Eq.~\eqref{nofieldderiv}, and the third term is negligible compared to the first term by Eq.\eqref{weakfield}.)  Notice the importance of keeping $v_\perp \ll 1$, since it appears multiplied by $\gamma \gg 1$.  

We may now solve for $v_\perp$ as
\begin{align}\label{vperp}
    v_{\perp}=\frac{\sqrt{\delta}}{\gamma},
\end{align}
where we define
\begin{align}\label{delta}
    \delta=\frac{E_0 \mathcal{E}}{E_0^2+B_0^2}.
\end{align}
Notice that $\delta$ satisfies $\delta \ll \gamma^2$ on account of $v_\perp \ll 1$ [Eq.~\eqref{approx}].  Although we assume that both $E_0 \ll \mathcal{E}$ and $B_0 \ll \mathcal{E}$, we cannot conclude that $\delta$ is large.

We now turn to the spatial portion of the LL equation \eqref{fi}.  Our goal is to express this equation in the Frenet-Serret frame of Sec.~\ref{sec:FS}.  Note, however, that the frame was defined relative to the PND $v_\pm = \ell(t)$, whereas the LL equation involves the full trajectory $\vec{v}(t)$.  In other words, the total derivatives $d/dt$ in Sec.~\ref{sec:FS} refer to a different curve from the total derivatives $d/dt$ in the present section.  However, by assumption these curves agree closely, and hence we may blur the distinction.  To see this explicitly, consider resolving a Frenet-Serret frame vector $\vec{e}$ using the total derivative of this section,
\begin{align}
    \frac{d\Vec{e}}{dt}  & = \Vec{v} \cdot \Vec{\nabla} \Vec{e} + \frac{\pd \Vec{e}}{\pd t} \\
    & \approx \Vec{l} \cdot \Vec{\nabla} \Vec{e} + \frac{\pd \Vec{e}}{\pd t} \\
    & \approx \Vec{l} \cdot \Vec{\nabla} \Vec{e}.
\end{align}
The first equation is true by definition, and the second follows from the assumption \eqref{approx} of near-PND motion.  The third then follows from the assumption \eqref{nofieldderiv} of a slowly varying field configuration, recalling that the frame vectors are determined by the field configuration.  However, $\Vec{l} \cdot \Vec{\nabla} \Vec{e}$ is simply the total derivative $d/dt$ of Sec.~\ref{sec:FS}, showing that the two are equivalent under our approximations.

We may therefore use the Frenet-Serret formulas \eqref{dl}-\eqref{dk} of Sec.~\ref{sec:FS} for the particle velocity at leading order.  Resolving the three-velocity into frame components yields
\begin{align}\label{lhs}
    \frac{d\Vec{v}}{dt}&=\left(\frac{d v_l}{dt}- v_n  \kappa\right)\Vec{l}+\left(\frac{dv_n}{dt}+v_l\kappa-\iota v_k \right)\Vec{n} \nonumber \\
    &\qquad +\left(\frac{dv_k}{dt}+\iota v_n \right)\Vec{k}.
\end{align}
Using Eqs.~\eqref{approx}, \eqref{dropvl} and \eqref{dropvn} [note that the first implies $v_\perp \ll 1$], we have
\begin{align}\label{aexp}
    \frac{d\Vec{v}}{dt}&=-v_n\kappa\Vec{l}+
    \left(\kappa-\iota v_k \right)\Vec{n}  + \left(\frac{dv_k}{dt} + \iota v_n \right)\Vec{k}.
\end{align}
It is necessary to retain the $dv_k/dt$ term at this stage because it will be the dominant term of the $\vec{k}$ component in cases where the torsion vanishes.

Eq.~\eqref{aexp} provides an approximation for the LHS of Eq.~\eqref{fi}.  For the first term of the RHS, we may similarly write $\Vec{F}$ [Eq.~\eqref{F}] as 
\begin{align}\label{Fexp}
    \Vec{F} =\pm \frac{E_0}{\gamma^2}(1+\delta)\Vec{l} \pm (B_0 v_k-E_0 v_n)\Vec{n}\mp(B_0 v_n+E_0 v_k)\Vec{k},
\end{align}
using the approximations \eqref{approx}.  Finally, the remaining terms on the RHS of \eqref{fi} contribute
\begin{align}\label{lastexp}
    \frac{1}{\mathcal{E}}\left[(\Vec{v} \times \Vec{E}) \times \Vec{E}+...\right] & = \mp\frac{E_0 }{\delta}\left( v_n \Vec{n} + v_k \Vec{k} \right) = \mp\frac{E_0}{\delta} \vec{v}_\perp.
\end{align}
The leading $\Vec{l}$ component is 
$E_0^2 v_\perp^2(\gamma^{-2}+v_\perp^2)/\mathcal{E}$, which is higher order and not written here.

Using Eqs.~\eqref{aexp}, \eqref{Fexp}, and \eqref{lastexp}, the resolution of Eq.~\eqref{fi} into the Frenet-Serret frame at leading order is 

\begin{align}
    - v_n \kappa & = \frac{|q| E_0}{\gamma^3 m}(1+\delta) \label{vl1} \\
    \kappa-\iota v_k & =\frac{|q|}{\gamma m}\left(B_0 v_k-E_0 v_n\frac{1+\delta}{\delta}\right) \label{vk1}\\
   \frac{dv_k}{dt} + \iota v_n & = -\frac{|q|}{\gamma m}\left(B_0 v_n+E_0 v_k\frac{1+\delta}{\delta}\right). \label{vn1}
\end{align}
 Eq.~\eqref{vl1} determines $v_n$ as
\begin{align}\label{vn}
    v_n=-\frac{|q| E_0}{\gamma^3 m \kappa}(1+\delta),
\end{align}
while Eq.~\eqref{vn1} tells us that
\begin{align}\label{vk}
    v_k=-\frac{\delta}{1+\delta}\left(\frac{B_0}{E_0}v_n+\left(\iota v_n + \frac{dv_k}{dt}\right)\frac{\gamma m}{|q| E_0}\right).
\end{align}
Eq.~\eqref{vk1} then  implies
\begin{align}\label{vn1.5}
    v_n=-\frac{\gamma m}{|q|}\frac{E_0\frac{1+\delta}{\delta}\kappa+(B_0+\frac{\gamma m}{|q|}\iota)\frac{dv_k}{dt}}{(B_0+\frac{\gamma m}{|q|}\iota)^2+E_0^2(\frac{1+\delta}{\delta})^2}.
\end{align}

Using Eqs.~\eqref{dropvk1} and \eqref{dropvk2}, Eqs.~\eqref{vk} and \eqref{vn1.5} become
\begin{align}\label{vk2}
    v_k & =-\frac{\delta}{1+\delta}\left(\frac{B_0}{E_0}+\iota \frac{\gamma m}{|q| E_0}\right)v_n. \\
    v_n & =-\frac{\gamma m E_0}{|q|}\frac{1+\delta}{\delta}\frac{\kappa}{( B_0+\frac{\gamma m}{|q|}\iota)^2+E_0^2\frac{(1+\delta)^2}{\delta^2}}.\label{vn2}
\end{align}

Setting Eq.~\eqref{vn2} equal to our previous expression \eqref{vn} for $v_n$, we derive 
\begin{align}
    \gamma^4
    &=\frac{3}{2}\frac{E_0}{|q| \kappa^2}\frac{( B_0+\frac{\gamma m}{|q|}\iota)^2+E_0^2\frac{(1+\delta)^2}{\delta^2}}{E_0^2+B_0^2}.\label{gammafull}
\end{align}
This is a quartic equation for the Lorentz factor $\gamma$.  Expanding out the terms, we find that the equation is equivalent to $f(\gamma)=0$ with 
\begin{align}
    f(\gamma) = \kappa^2\gamma^4-\delta\iota^2\gamma^2-2\iota \delta \frac{|q| B_0}{m} \gamma-\frac{3}{2}\frac{E_0}{|q|},\label{f}
\end{align}
where we have used $E_0 \ll \mathcal{E}$ and $B_0 \ll \mathcal{E}$ [Eq.~\eqref{weakfield}].

The possible equilibrium Lorentz factors are the positive, real roots of $f(\gamma)$.  Since $f(0)<0$ and $f(\infty)=\infty$, there must be at least one positive, real root.  If $\iota>0$ then there is a single such root, since the last three terms in \eqref{weakfield} comprise a monotonically decreasing function on $\gamma>0$.  However, if $\iota < 0$, there can be up to three positive, real roots, indicating up to three separate equilibria.  It is not clear at this stage whether some selection rule distinguishes a single, physical equilibrium, or whether all three are allowed.\footnote{An example of a three-root case that respects the assumptions of the derivation is $E_0=10^{-8}\mathcal{E}$, $B_0=10^{-6}\mathcal{E}$,  $\kappa=7 \times 10^{-14}\mathcal{R}^{-1}$, and $\iota=-10^3\kappa$. Notice in particular that $|\iota| \gg \kappa$; all examples we have found share this feature.}  We hope to address this issue in a future paper in this series.

Once $\gamma$ is determined, the other subleading velocity components $v_n$ and $v_k$ are given directly by Eqs.~\eqref{vn} and \eqref{vk2}.  This completes the equilibrium description.

\subsection{Negligible torsion case}\label{sec:torsion}

The equilibrium description simplifies significantly if the torsion can be neglected.  This occurs if either $E_0$ or $B_0$ is sufficiently large,\footnote{If $\iota \ll |q|B_0/(m \gamma)$, we see from Eq.~\eqref{gammafull} that $\iota$ disappears from the description.  If this condition is not satisfied, but instead $\iota \ll |q|E_0/(m \gamma)$, then we must have $B_0 \ll E_0$, meaning $\delta \approx \mathcal{E}/E_0$ and it can be seen from \eqref{f} that the second and third terms are negligible compared to the last term}
\begin{align}\label{iotasmall}
    |\iota| \ll \frac{|q|}{m \gamma}\textrm{Max}\{E_0,B_0\}.
\end{align}

Then the condition $f(\gamma)=0$ \eqref{f} has the unique solution 
\begin{align}\label{gammaalmostfinal}
    \gamma^4=\frac{3}{2}\frac{E_0}{|q| \kappa^2}.
\end{align}
This is Gruzinov's result \eqref{eq:gruzi}.  Plugging in to Eqs.~\eqref{vn} and \eqref{vk} gives the perpendicular velocity as
\begin{align}
    v_n&=-\frac{1+\delta}{\gamma}\sqrt{\frac{E_0}{\mathcal{E}}} \label{vnalmostfinal}\\
    v_k&=\frac{\delta}{\gamma}\left(\frac{B_0}{E_0}+\frac{\gamma m}{|q| E_0}\iota\right)\sqrt{\frac{E_0}{\mathcal{E}}}.\label{vkalmostfinal}
\end{align}

We have retained the torsion in Eq.~\eqref{vkalmostfinal} because we only assume that $m\gamma\iota/q$ is small compared to $E_0$ \textit{or} $B_0$ [Eq.~\eqref{iotasmall}].  However, consider each case separately.  If it is small compared to $B_0$, then the torsion term is negligible in \eqref{vkalmostfinal}.  If it is small compared to $E_0$ but not $B_0$, then we have $E_0 \gg B_0$ and $\delta \gg 1$, which makes \textit{both} terms negligible in \eqref{vkalmostfinal}, i.e., we have $v_k \ll v_n$.  In both cases the torsion is negligible, so we may in fact drop it from our final answer.  Introducing the radius of curvature $R=1/\kappa$, we present the final results as 
\begin{align}
    \gamma^4&=\frac{9}{4}\left(\frac{ R}{\mathcal{R}}\right)^2\frac{E_0}{\mathcal{E}}\label{gammafinal} \\
    v_n&=-\frac{1+\delta}{\gamma}\sqrt{\frac{E_0}{\mathcal{E}}} \label{vnfinal}\\
    v_k&=\frac{\delta}{\gamma}\frac{B_0}{E_0}\sqrt{\frac{E_0}{\mathcal{E}}},\label{vkfinal}
\end{align}
where $\delta$ was given in Eq.~\eqref{delta}.

\subsection{Consistency Conditions for Equilibrium}\label{sec:conditions}

We have now obtained explicit expressions \eqref{gammafinal}-\eqref{vkfinal} for the equilibrium velocity using the nine assumptions in Sec.~\eqref{sec:assumptions} together with the additional assumption \eqref{iotasmall}.  The velocity expressions may now be plugged into the assumptions to determine consistency conditions on the field configuration.  The resulting conditions are not all independent, and we will choose the following five
\begin{align}
    \gamma & \gg 1 \\
    v_\perp & \ll 1 \\ 
    |\iota| & \ll \frac{|q|}{m\gamma}\textrm{Max}\{E_0,B_0\} \\ 
    m \left|\frac{d\gamma}{dt}\right| & \ll |q| E_0, \\
    T & \gg L, \label{collected}
\end{align}

which in fact imply the remaining conditions (App.~\ref{sec:cleanup}).  Using Eqs.~\eqref{gammafinal}--\eqref{vkfinal}, together with \eqref{scales}, these conditions respectively become
\begin{align}
    C_1 & = \frac{\mathcal{R}}{R} \sqrt{\frac{\mathcal{E}}{E_0}}  \ll 1 \label{C1} \\
    C_2 & = \frac{\mathcal{R}}{R} \sqrt{\frac{\mathcal{E}}{E_0}}\frac{E_0\mathcal{E}}{E_0^2+B_0^2} \ll 1 \label{C2} \\
    C_3 & = |\iota| \sqrt{R \mathcal{R}} \left(\frac{E_0}{\mathcal{E}}\right)^{\frac{1}{4}} \frac{\mathcal{E}}{\textrm{Max}\{E_0,B_0\}} \ll 1 \label{C3} \\
    C_4 & = \frac{\sqrt{\mathcal{R}R}}{L}\left(\frac{\mathcal{E}}{E_0}\right)^{3/4}  \ll 1, \label{C4}\\
    C_5 &= \frac{L}{T} \ll 1 \label{C5}
\end{align}

where $\mathcal{R}$ and $\mathcal{E}$ are the length and field scales defined by the charge and mass [Eq.~\eqref{scales}], while $L$ and $T$ are the typical length and time scales in the lab frame.  Here we have defined five dimensionless quantities $C_i$, which all must be small for the equilibrium to be possible.  Note that only the length scale appears in \eqref{C4} on account of the assumption \eqref{C5}.

Notice that the conditions \eqref{C1} and \eqref{C4} are always violated in the limit $E_0 \to 0$.  In particular, our results do not apply to purely magnetic fields ($E_0=0$) or to null electromagnetic fields ($|\vec{E}|=|\vec{B}|$ and $\vec{E}\cdot\vec{B}=0$, implying $E_0=B_0=0$) such as plane waves.  (Plane waves also violate the condition \eqref{C5}.)  However, Refs.~\cite{Gonoskov:2017lyz,Samsonov:2018skj,Gonoskov:2021hwf,Ekman:2021vwg,Samsonov2022} demonstrate circumstances in which particles still follow the principal null directions in electromagnetic fields of relevance to laser-plasma interaction.  Further work is necessary to connect this phenomenon to the one studied here.

The conditions \eqref{C1}--\eqref{C4} involve the local electromagnetic field strength only through the invariants $E_0$ and $B_0$, which are independent of the choice of frame.  However, they involve its \textit{scale of variation} through a number of non-invariant quantities: the curvature radius $R$ and torsion $\iota$ (which are defined relative to spatial projections of the principal null directions in a particular frame), as well as the more basic field variation parameters $L$ and $T$.  Thus our conditions \eqref{C1}--\eqref{C4} are  not Lorentz-invariant, but rather are tied to a specific frame where the motion is ultrarelativistic.  

In other words, our assumption on the field configuration is that \textit{there exists a frame} where the $C_i$ are small.  In the typical applications of pulsars and lasers, there is always a natural frame to consider (in which the star or apparatus is at rest), but as a theoretical issue this outcome is somewhat unsatisfying.  It may be that restoring consideration of whether the particle actually enters the regime restores Lorentz invariance (as explored, for example, in Ref.~\cite{Jacobson:2015cia}), or it may be that this ultrarelativistic approximation is necessarily non-invariant, since it requires a preferred frame in which the motion is ultrarelativistic.

To summarize, we have shown that Aristotelian equilibrium can occur when the parameters $C_i$ are small, and that the velocity in this case is given by Eqs.~\eqref{gammafinal}-\eqref{vkfinal}.  If the torsion condition $C_3 \ll 1$ is not satisfied, the Aristotelian equilibrium can still occur, but the velocity must be determined by solving a quartic equation \eqref{gammafull}. However, in this case we do not obtain simple conditions (analogous to $C_1$, $C_2$, and $C_4$) for the consistency of the equilibrium.

\begin{figure*}
    \centering
    \includegraphics[width=.45\textwidth]{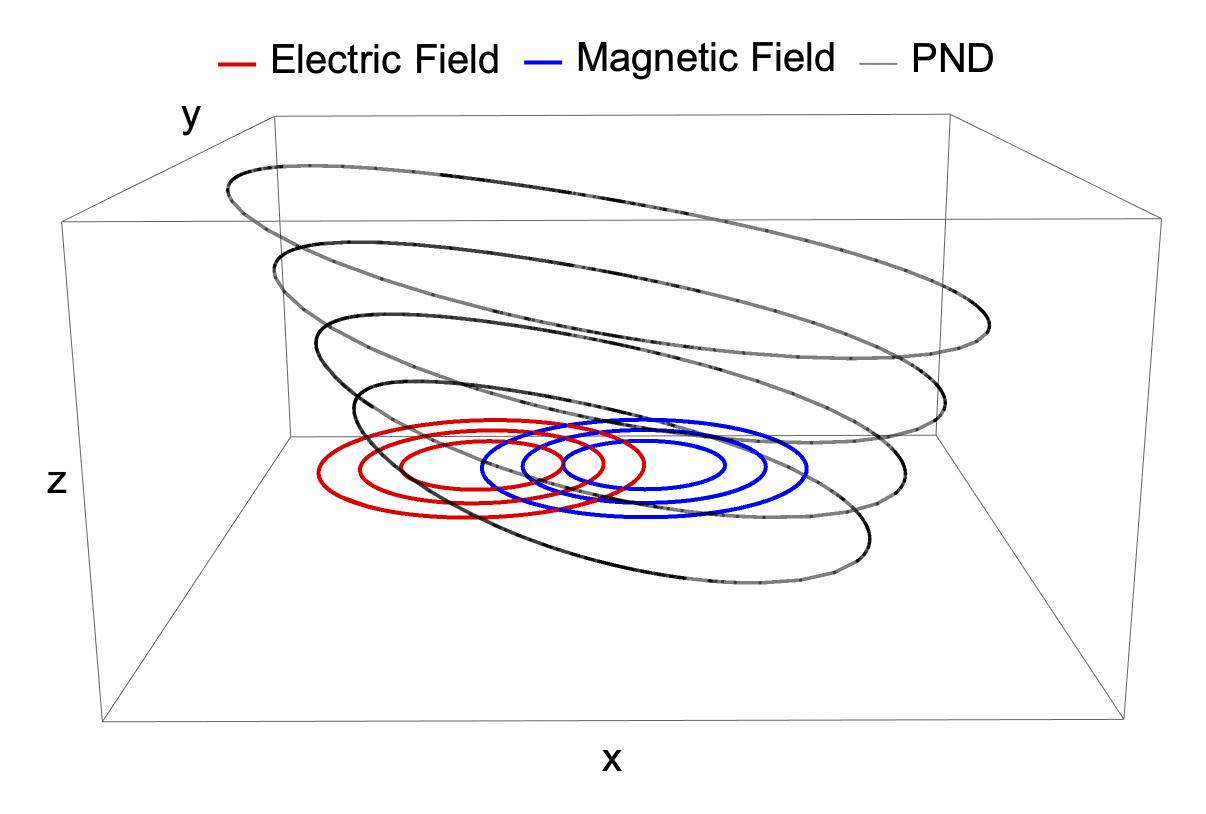}
    \includegraphics[width=.45\textwidth]{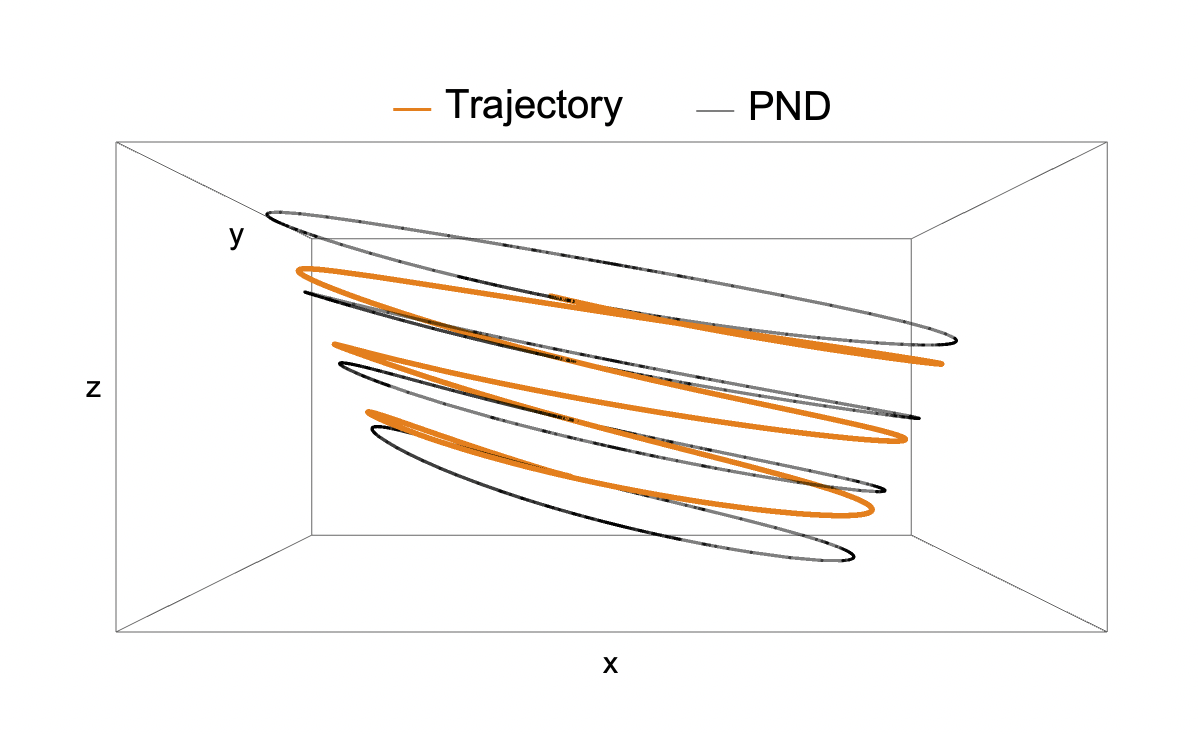}
    \caption{A simple example of Aristotelian equilibrium.  The field configuration consists of offset circular magnetic and electric fields, with the magnetic field twice as strong as the electric field (see Sec.~\ref{sec:numerical} for details).  The spatial projections of the PNDs are closed tilted ovals, shown in black.  The right panel shows the  trajectory of a charged particle determined by numerically solving the LL equation \eqref{LL}.  The trajectory remains locally tangent to a PND as the particle slowly drifts from one to another.}
    \label{fig:3dplots}
\end{figure*}

\section{Numerical Comparison}\label{sec:numerical}

We have performed numerical simulations of the full LL equation \eqref{LL} using the implicit Runge-Kutta-Nystr\"{o}m method described in Ref.~\cite{numeric1}, augmented with an improved iteration method to handle the large Lorentz factors present for Aristotelian motion.  In this section we report results showing the accuracy of the analytic approximation in one example where the assumptions \eqref{C1}-\eqref{C4} are satisfied.  In a follow-up paper \cite{inprep} we will present details of this code and use it to more fully explore the parameter space of Aristotelian motion, together with the important question of the approach to equilibrium.

Our example is artificial and chosen for simplicity: we consider electric and magnetic fields that are circular, with constant field strength, but with offset centers.  That is, we let $\Vec{B}=B \hat{\phi}$, where $B$ is constant and $\hat{\phi}$ is the unit vector circulating around some choice of $z$ axis.  We similarly let $\vec{E}=E \hat{\phi}'$, where $E$ is constant and $\hat{\phi}'$ is the unit vector circulating around some different axis $z'$.  The $z$ and $z'$ axes are parallel, separated by a distance $D$.  The spatial principal null directions (integral curves of $\Vec{v}_\pm$) are closed ovals tilted ovals, as shown in Fig.~\ref{fig:3dplots}.

We choose $E=3.13\times10^{-7}\mathcal{E}$, $B=2 E$, and $D=2.13\times10^{11}\mathcal{R}$, making the dimensionless parameters \eqref{C1}--\eqref{C4} of magnitude $C_1 \approx 10^{-9}$, $C_2 \approx 10^{-3}$, $C_3 \approx 10^{-5}$, and $C_4 \approx 10^{-2}$.  We find that the particle enters the Aristotelian equilibrium within a timescale of order $\sim 10 m/(q E)$, irrespective of initial conditions, after which the numerical trajectories agree closely with the analytic predictions.

\begin{figure}
    \centering
    \includegraphics[width=.45\textwidth]{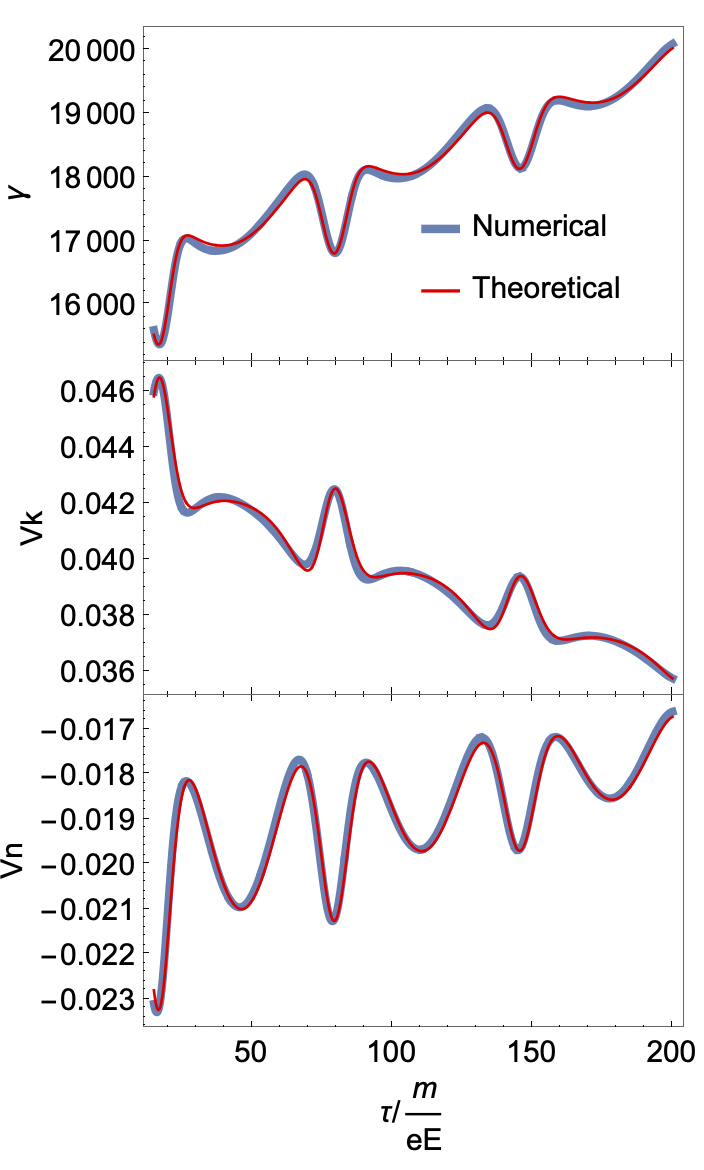}\\
    \caption{Agreement of the analytic predictions \eqref{gammafinal}--\eqref{vkfinal} with numerical simulation of the LL Equation \eqref{LL}, for the example of Fig.~\ref{fig:3dplots}.  The maximum difference in $\gamma$ is 0.5\%, in $v_n$ is 3\%, and in $v_k$ is 1.5\%.  This level of agreement is commensurate with the sizes of the dimensionless parameters $C_i$ controlling the validity of the approximation (see the third paragraph of Sec.~\ref{sec:numerical}). 
    }
    \label{fig:2dplots}
\end{figure}

We present one example in detail  (Figs.~\ref{fig:3dplots} and \ref{fig:2dplots}).  The particle starts at position $(0,1.5 D,0)$ with initial velocity pointing in the $-x$ direction and a $\gamma$ factor of 500.  The plots begin at proper time $15\frac{m}{q E}$ when the Lorentz force power and self-force power agree to within $5\%$, signaling entry into the Aristotelian equilibrium.  The particle remains nearly tangent to the PNDs, while drifting slowly from one to the other (Fig.~\ref{fig:3dplots}).  The analytic predictions  \eqref{gammafinal}--\eqref{vkfinal} for the Lorentz factor and drift velocities agree with the full numerical solution (Fig.~\ref{fig:2dplots}).

\section{Outlook}

 In this paper we have studied the Aristotelian equilibrium of ultrarelativistic charged particles, deriving the detailed description \eqref{gammafinal}--\eqref{vkfinal} of the particle velocity as well as the associated  conditions \eqref{C1}--\eqref{C4} on the field configuration for this equilibrium to be possible.  We have also demonstrated numerically that the equilibrium occurs in a simple example, and quantitatively checked the analytic predictions in this case. 

A number of important questions remain to be addressed.  First, are the conditions \eqref{C1}--\eqref{C4} \textit{sufficient} to enter equilibrium, or are additional conditions required?  And how quickly does this happen?  Second, can a simple description be given when the torsion is not negligible [Sec.~\ref{sec:torsion}]?  Third, does the curvature of spacetime modify the dynamics in an important way when the spacetime curvature radius is comparable to that of the electromagnetic field (as occurs, for example, in pulsars)?  And finally---and perhaps most importantly---does the Aristotelian equilibrium actually occur near pulsars, near black holes, or in man-made laser plasmas, and can we understand the resulting phenomenology?  We plan to address these issues in future papers in the series.

\section*{Acknowledgements}

This work was supported in part by NSF grant PHY–1752809 as well as NSF grants PHY-1912619 and PHY-2145421 to the University of Arizona.

\appendix

\section{Arguments for Aristotelian motion}\label{sec:previous}
In our treatment of Aristotelian motion, the leading-order motion along a PND is an assumption, rather than a result.  While it would be desirable to instead derive the PND motion from some more fundamental assumption, we have thus far been unsuccessful.  In this section we review previous arguments for PND motion, found in four separate, independently conceived papers.  While strongly suggestive, and extremely useful for building intuition for the phenomenon, none of these arguments rises to the standard of rigor we would prefer to adopt.

\subsection{Herold et. al, 1985 \cite{Herold1985}}

Herold et. al \cite{Herold1985} (see also \cite{BFink1989}) presented some basic arguments for Aristotelian motion in the context of pulsar magnetospheres.  In the language of this paper, their approach is as follows.

First recall that the middle two terms of the energy equation \eqref{f0} are negligible compared to the first term and can be dropped in a study of Aristotelian motion (see discussion below \eqref{f0}). Using the assumption $\gamma \gg 1$, the last term is proportional to $\vec{F}^2$ [see Eq.~\eqref{F}], and Eq.~\eqref{f0} becomes
\begin{align}
        \frac{d\gamma}{dt}&=\frac{q}{m}\left\{ \Vec{E}\cdot\Vec{v} \mp\frac{\gamma^2}{\mathcal{E}}\vec{F}^2\right\},
\end{align}
where $\pm$ is the sign of $q$.  For the two terms to be the same order of magnitude (reflecting Aristotelian equilibrium), we must have
\begin{align}
    \frac{\vec{F}^2}{|\vec{E}\cdot \vec{v}|^2} \approx \frac{\mathcal{E}}{\gamma^2|\vec{E}\cdot\vec{v}|}.\label{moo}
\end{align}
Ref.~\cite{Herold1985} then argues that the right-hand-side will be small for typical pulsar parameters (their dimensionless damping constant $D_0$ is large).  They thus conclude that $\vec{F}$ is much smaller than the typical field scale (here represented by $\vec{E}\cdot \vec{v}$), and hence can be approximated as zero.  Solving the condition $\vec{F} = 0$ gives rise to motion along a PND  [see Eqs.~\eqref{pnd1} and \eqref{pnd2}].

The main drawback of this approach is that it makes an assumption on the specific size of $\gamma$ relative to other parameters in the problem---namely that the RHS of \eqref{moo} is small.  We find it more natural to derive the size of $\gamma$ [and, in fact, the full formula \eqref{gammafinal}] from the assumption of motion along a PND.  We emphasize, however, that both approaches require an assumption about the particle's state of motion.  It would be preferable to find conditions purely on the electromagnetic field and intrinsic particle properties (mass and charge), such that Aristotelian motion will inevitably occur.

Refs.~\cite{Herold1985,BFink1989} also presented numerical simulations of a truncated LL equation, 
\begin{align}
    \frac{d\gamma}{dt}&=\frac{q}{m}\left\{ \Vec{E}\cdot\Vec{v}\mp\frac{\gamma^2}{\mathcal{E}}\left[(\Vec{E}+\Vec{v}\times\Vec{B})^2-(\Vec{E}\cdot\Vec{v})^2\right]\right\} \label{f02} \\
    \frac{d\Vec{v}}{dt}&=\frac{q}{m\gamma} \vec{F}. \label{fi2}
\end{align}
These equations arise after dropping the second and third terms in each of Eqs.~\eqref{f0} and \eqref{fi}.  While this approximation is valid for \eqref{f0} (as already discussed), it is not valid for \eqref{fi}.  The last term in \eqref{fi} would be small compared to the first if $\vec{F}$ were of its naive size $\sim \vec{E}$, but the Aristotelian regime is precisely where $|\vec{F}| \ll |\vec{E}|$ [see discussion below Eq.~\eqref{moo}].  In fact, the third term in \eqref{fi} is comparable to the first and cannot be dropped; its contribution to the equilibrium appears in Eq.~\eqref{lastexp} and propagates through the remainder of the derivation.  If this contribution were left out, incorrect formulas for the subleading velocity components \eqref{gammafinal}--\eqref{vkfinal} would be obtained, which do not match numerical simulations of the full LL equation.

\subsection{Mestel et al., 1985 \cite{mestel1985}}

Mestel et al. \cite{mestel1985} argued for Aristotelian motion in the context of pulsar magnetospheres.  They suggest that at large $\gamma$ two properties should emerge: (1) the self-force is just $-\mathcal{P}\vec{v}$, where $\mathcal{P}$ is the instantaneous power radiated (relativistic Larmour formula), and (2) inertial terms are negligible.  This results in the equation\footnote{Eqs.~(6.20) and (6.21) of Ref.~\cite{mestel1985} appear to have typographical errors; the Lorentz force term should not have a minus sign.}
\begin{align}\label{Mbalance}
    q(\vec{E}+\vec{v}\times\vec{B})-\mathcal{P}\vec{v}=0,
\end{align}
which is precisely the eigenvalue equation for the PNDs [see Eq.~\eqref{pnd1}].  The authors of \cite{mestel1985} thus conclude that the motion is along the PND.

The main drawback of this approach is that the assumptions (1) and (2) leading to \eqref{Mbalance} are heuristic, and not derived from the fundamental LL description of the problem.  However, we find this approach to be quite valuable for intuition.  At high Lorentz factor, it is rather natural to expect the mass to be negligible and for the self-force to be pure radiation damping, even if it is not clear precisely how (or whether!) these features emerge from the LL equation \eqref{fi}.  There is also a nice connection with the guiding center theory that governs motion in strong magnetic fields.  If one ``turns off'' radiation reaction by letting $\mathcal{P}=0$, then Eq.~\eqref{Mbalance} becomes the guiding center condition $\vec{E} + \vec{v} \times \vec{B}=0$.  We can regard the guiding center and Aristotelian regimes as  distinct zero-inertia limits of charged particle motion, depending on the importance of radiation reaction.

\subsection{Gruzinov, 2013 \cite{Gruzinov:2013pva,gruzinov2013b}}

Gruzinov discussed the Aristotelian regime in \cite{Gruzinov:2013pva} as part of a model for pulsar magnetospheres and gave further details and derivations in \cite{gruzinov2013b}.  He simply assumed that particles follow the PNDs, reasoning that they ought to move along the field direction in the special family of frames where $\vec{E}$ and $\vec{B}$ are parallel.  He then argued that the terminal Lorentz factor should be determined by setting the power delivered by Lorentz force equal to the energy lost due to curvature radiation.  As described in the text above Eq.~\eqref{eq:gruzi}, this leads directly to the formula \eqref{gammafinal} [or \eqref{eq:gruzi}] for the Lorentz factor.

Our work provides a derivation of Gruzinov's result from the fundamental dynamical equation (the LL equation), which reveals four specific conditions \eqref{C1}--\eqref{C4} that are required for the result to hold.  We also obtain the full subleading description, finding the drift velocities \eqref{vnfinal} and \eqref{vkfinal} in addition to the Lorentz factor \eqref{gammafinal}.

\subsection{Laser-plasma community, 2018 \cite{Gonoskov:2017lyz,Samsonov:2018skj}}

Gonoskov and Marklund \cite{Gonoskov:2017lyz} (see also \cite{Gonoskov:2021hwf}) and Samsonov et. al \cite{Samsonov:2018skj} (see also \cite{Samsonov2022}) analyzed Aristotelian motion in the context of laser-plasma physics, using the name ``radiation free directions'' for the PNDs.  While the basic phenomenon of motion along PNDs also occurs in this context, the details are quite different since these authors focus on plane wave configurations (excluded in our approach) and allow more general forms of the radiation reaction force, including quantum radiation.  It would be interesting to explore the connections further.

In Ref.~\cite{Gonoskov:2017lyz}, the analytical arguments for motion along a PND involve a uniform field configuration.  If we restrict to the LL equation, the uniform field case can be understood completely using the general analytic solution \cite{Heintzmann1973}.  This solution demonstrates that, irrespective of initial conditions, the Lorentz factor of the particle approaches $\infty$ at late times, with the velocity approaching the field direction (PND).  This is crucially different from the Aristotelian equilibrium we study, where the field radius of curvature $R$ limits the ultimate Lorentz factor.  The uniform field case represents the limit $R \to \infty$, where there is no radiation reaction force and the equilibrium we study cannot occur.

\section{Analysis of weaker conditions}\label{sec:cleanup}

In Sec.~\ref{sec:conditions} it was claimed that, once the equilibrium formulas for velocity are used, the conditions $C_1$--$C_5$ (Eqs.~\eqref{C1}--\eqref{C5}) are sufficient to guarantee that all other assumptions of the derivation are satisfied.  To justify this claim, we must check that Eqs.~\eqref{C1}--\eqref{C5} imply $L \gg \gamma \mathcal{R}$ \eqref{nofieldderiv} as well Eqs.~\eqref{dropvl}--\eqref{dropvk2}.  We will consider each equation in turn.

First consider $L \gg \gamma \mathcal{R}$.  Using Eq.~\eqref{gammafinal}, this becomes
\begin{align}\label{oink}
    \frac{\sqrt{\mathcal{R}R}}{L}\left(\frac{E_0}{\mathcal{E}}\right)^{\frac{1}{4}} \ll 1,
\end{align}
which is implied by $C_4 \ll 1$ \eqref{C4} since $E_0 \ll \mathcal{E}$ \eqref{weakfield}.

Next consider Eq.~\eqref{dropvl}.  First note that $v_l$ is related to the Lorentz factor $\gamma$ by
\begin{align}
    v_l=\sqrt{1-\frac{1}{\gamma^2}-\frac{\delta}{\gamma^2}},
\end{align}
where Eq.~\eqref{vperp} was used.  The time derivative is thus
\begin{align}
    \frac{dv_l}{dt}=\frac{1}{v_l}\left(\frac{1}{\gamma^3}\frac{d\gamma}{dt}+\frac{\delta}{\gamma^3}\frac{d\gamma}{dt}-\frac{1}{2}\frac{1}{\gamma^2}\frac{d\delta}{dt}\right).
\end{align}
Since $\gamma$ and $\delta$ are completely determined by the field configuration in equilibrium, they inherit its scale of variation as
\begin{align}\label{woof}
    \frac{d\gamma}{dt}\lesssim \frac{\gamma}{L}, \qquad \frac{d\delta}{dt} \lesssim \frac{\delta}{L}.
\end{align}
(We need not consider the timescale $T$ on account of the assumption $T \gg L$, c.f. Eq.~\eqref{C5} above.)  Using $v_l \approx 1$, we thus have
\begin{align}
    \frac{dv_l}{dt} \lesssim \frac{1+\frac{\delta}{2}}{\gamma^2}\frac{1}{L}.
\end{align}
Using this expression together with \eqref{vnfinal} and \eqref{gammafinal} as well as $\kappa=1/R$ and $v_\ell\approx 1$, Eq.~\eqref{dropvl} becomes
\begin{align}
    \frac{\sqrt{\mathcal{R}R}}{L}\left(\frac{\mathcal{E}}{E_0}\right)^{\frac{3}{4}}\frac{1+\delta/2}{1+\delta} \ll 1,
\end{align}
which is implied by $C_4 \ll 1$ \eqref{C4}. 

Next consider Eq.~\eqref{dropvn}.  Using $dv_n/dt \lesssim v_n/L$ [similarly to Eq.~\eqref{woof}] with Eq.~\eqref{vnfinal} as well as $\kappa=1/R$ and $v_l\approx 1$, we find that \eqref{dropvn} becomes
\begin{align}
    \frac{\sqrt{\mathcal{R}R}}{L}\left(\frac{E_0}{\mathcal{E}}\right)^{\frac{1}{4}}(1+\delta) \ll 1.\label{cluck}
\end{align}
If $\delta \lesssim 1$, then this condition is equivalent to \eqref{oink}, which we have already seen is implied by $C_4 \ll 1$.  If instead $\delta \gg 1$, then using the definition \eqref{delta} of $\delta$, Eq.~\eqref{cluck} becomes
\begin{align}
    \frac{\sqrt{\mathcal{R}R}}{L}\left(\frac{\mathcal{E}}{E_0}\right)^{\frac{3}{4}} \frac{E_0^2}{E_0^2+B_0^2} \ll 1,
\end{align}
which is also implied by $C_4 \ll 1$ \eqref{C4}.  Thus Eq.~\eqref{dropvn} is indeed satisfied.

Next consider Eq.~\eqref{dropvk1}. Using $d v_k/dt \lesssim v_k/L$ [similarly to \eqref{woof}], this condition becomes
\begin{align}\label{alsogamma}
\frac{\gamma m}{|q| E_0 L}\frac{\delta}{1+\delta} \ll 1,
\end{align}
which is the same as Eq.~\eqref{gammadot} (equivalently $C_4 \ll 1$ \eqref{C4}), using Eq.~\eqref{woof}.

Finally consider Eq.~\eqref{dropvk2}.  Using Eq.~\eqref{vperp}, this condition becomes
\begin{align}\label{condvk2}
    \left|\frac{dv_k}{dt}\right| \ll \frac{ E_0}{\left|B_0+\frac{\gamma m}{|q|}\iota\right|}\frac{1+\delta}{\delta}\kappa.
\end{align}
From Eqs.~\eqref{vn} and \eqref{vk2}, we know that 
\begin{align}
    v_k=\frac{|q|}{\gamma^3 m \kappa}\left(B_0+\frac{\gamma m}{|q|}\iota\right)\delta,
\end{align}
meaning that Eq.~\eqref{condvk2} becomes
\begin{align}
    \left|\frac{dv_k}{dt}\right| \ll \frac{1+\delta}{\left|v_k\right| \gamma^3}\frac{|q| E_0}{m},
\end{align}
or again using $dv_k/dt \lesssim v_k/L$,
\begin{align}\label{condvk}
    \frac{\gamma m}{|q|E_0 L} \frac{\gamma^2 v_k^2}{1+\delta} \ll 1.
\end{align}
This equation is equivalent to \eqref{dropvk2} under our assumptions.  To prove that it is satisfied, note that the first factor is small by \eqref{alsogamma}, while the second satisfies
\begin{align}
    \frac{\gamma^2 v_k^2}{1+\delta} & < \frac{\gamma^2 v_{\perp}^2}{1+\delta}=\frac{\delta}{1+\delta} < 1,
\end{align}
noting that $v_\perp^2 = v_n^2 + v_k^2$ and using Eq.~\eqref{vperp}.

\bibliographystyle{utphys}
\bibliography{APSdraft}

\end{document}